\documentclass[prl,amsmath,amssymb,twocolumn]{revtex4}

\usepackage{graphicx}
\usepackage{dcolumn}
\usepackage{bm}
\usepackage{epstopdf}
\usepackage{epsfig}
\usepackage{multirow}

\usepackage{color}
\usepackage{ulem}

\begin{document}

\title{A high-sensitivity gate-based charge sensor in silicon}

\author{M.~F.~Gonzalez-Zalba}
\email{mg507@cam.ac.uk}
\thanks{These authors contributed equally to this work}
\affiliation{Hitachi Cambridge Laboratory, J. J. Thomson Avenue, Cambridge CB3 0HE, United Kingdom}
\author{S. Barraud}
\affiliation{SPSMS, UMR-E CEA / UJF-Grenoble 1, INAC, 17 rue des Martyrs, 38054 Grenoble, France}
\author{A.~J.~Ferguson}
\affiliation{Cavendish Laboratory, University of Cambridge, CB3 0HE, United Kingdom}
\author{A.~C.~Betz}
\email{ab2106@cam.ac.uk}
\thanks{These authors contributed equally to this work}
\affiliation{Hitachi Cambridge Laboratory, J. J. Thomson Avenue, Cambridge CB3 0HE, United Kingdom}
\date{\today}

\maketitle

{\bfseries The implementation of a quantum computer requires a qubit-specific measurement capability to read-out the final state of a quantum system. The model of spin dependent tunneling followed by charge readout has been highly successful in enabling spin qubit experiments in all-electrical, semiconductor based quantum computing. As experiments grow more sophisticated, and head towards multiple qubit architectures that enable small scale computation, it becomes important to consider the charge read-out overhead. With this in mind, Reilly et al. demonstrated a gate readout scheme in a GaAs double quantum dot that removed the need for an external charge sensor. This readout, which achieved sensitivities of order me/$\sqrt(Hz)$, was enabled by using a resonant circuit to probe the complex radio-frequency polarisability of the double quantum dot. However, the ultimate performance of this technology and the noise sources that limit it remain to be determined. Here, we investigate a gate-based readout scheme using a radio-frequency resonant circuit strongly coupled to a double quantum at the corner states of a silicon nanowire transistor. We find a significantly improved charge sensitivity of 37 $\mu$e/$\sqrt(Hz)$. By solving the dynamical master equation of the fast-driven electronic transitions we quantify the noise spectral density and determine the ultimate charge and phase sensitivity of gate-based read-out. We find comparable performance to conventional charge sensors and fundamental limits of order ne/$\sqrt(Hz)$ and $\mu$rad/$\sqrt(Hz)$, with the gate-based sensor improving on standard detection for certain device parameters. Our results show that, especially in state-of-the-art silicon qubit architectures, charge detection by probing the complex polarisability has advantages in terms of reducing the readout overhead but also in terms of the absolute charge sensitivity.}

High-fidelity quantum state readout is a requirement for the successful implementation of a quantum computer. Traditionally, in semiconductor quantum dots this has been achieved by charge sensing techniques that require a separate electrometer. The most sensitive of these charge sensors are the radio-frequency quantum point contact (rf-QPC)~\cite{vanWees1988,Reilly2007,Cassidy2007} and the radio-frequency single electron transistor (rf-SET)~\cite{Kastner1992,Schoelkopf1998, Lu2003}, where the respective device is embedded in an impedance matching circuit. Charge sensitivities as good as 100~$\mu$e/$\sqrt{Hz}$~\cite{Mason2010} for the rf-QPC and $1\,(0.9)\,\mu$e/$\sqrt{Hz}$ for the normal (superconducting) rf-SET ~\cite{Aassime2001a,Brenning2006} have been achieved with Megahertz bandwidth resolution. 

With the advent of circuit quantum electrodynamics an alternative method to readout the state of a quantum system became apparent: A microwave resonator's self-resonance is modified by the state-dependent polarisability of mesoscopic systems connected to it. Embedding the qubit in a high quality cavity resonator allows to readout the qubit state via the dispersive and dissipative interaction with the cavity. This technique has been demonstrated for superconducting qubits~\cite{Wallraff2004,Blais2004,Niemczyk2010,Barends2014} and semiconductor charge~\cite{Frey2012a,Viennot2014} and spin qubits~\cite{Petersson2012,Kubo2010}.\newline
In the same spirit, resonant LC circuits have been coupled directly to the ohmic contacts of double quantum dots to readout their charge and spin state~\cite{Petersson2010, Schroer2012}, and to determine their complex admittance~\cite{Chorley2012}. Recently, it has been demonstrated that the gates defining these mesoscopic systems can also act as fast and sensitive readout elements~\cite{Colless2013}.

While the high-frequency polarisability changes are well understood in terms of Sisyphus resistance~\cite{Persson2010}, and state-dependent quantum or tunnelling capacitance~\cite{Ashoori1992,Persson2010,Ciccarelli2011,Lambert2014} the sensitivity of this approach seemed to be experimentally limited to the me/$\sqrt{Hz}$ range for GaAs technology~\cite{Petersson2010,Johansson2006}. Moreover, the cyclostationary noise~\cite{Roschier2004} inherent to this technique has not been studied yet and needs to be addressed in order to determine the fundamental limit of charge and phase sensitivity of fast gate-based detection.

In the case of silicon technology, only standard QPC and SET detectors have been used to readout coherent properties of silicon quantum dots~\cite{Maune2012,Shi2014}, and dopant electron and nuclear spins~\cite{Morello2010, Pla2012, Pla2013,Yin2013,Dupont-Ferrier2013}. Although fast detectors have been demonstrated~\cite{Angus2008,Manoharan2008,Nishiguchi2013} little progress has been made towards gate-based sensing~\cite{Villis2011}, despite the fact that silicon nanowire transistors present an excellent platform for gate detection.

In this Article we demonstrate and benchmark a high-sensitivity gate-based charge sensor and develop a model to calculate the ultimate sensitivity of this type of detector. The sensor is implemented on the gate electrode of a narrow tri-gate nanowire field-effect transistor (NWFET). By coupling the gate to a radio-frequency resonant circuit we probe the charge state of a few-electron double quantum dot in silicon.  Due to the strong capacitive coupling between the resonator and the quantum dots ($\alpha=0.92$~\footnote{The gate coupling is given by $\alpha=C_g/C_{\Sigma}$, where $C_g$ is the gate and $C_{\Sigma}$ the total capacitance, respectively}) we obtain a charge sensitivity of $\delta q=37\,\mu e/\sqrt{Hz}$, a two-order magnitude improvement to previously reported GaAs sensors~\cite{Colless2013}. We demonstrate that the measured sensitivity is experimentally limited by the cryogenic amplifier noise and theoretically by the Sisyphus noise predicting an ultimate sensitivity that could outperform rf-SETs.

\begin{figure*}[htbp]
	\centering
		\includegraphics{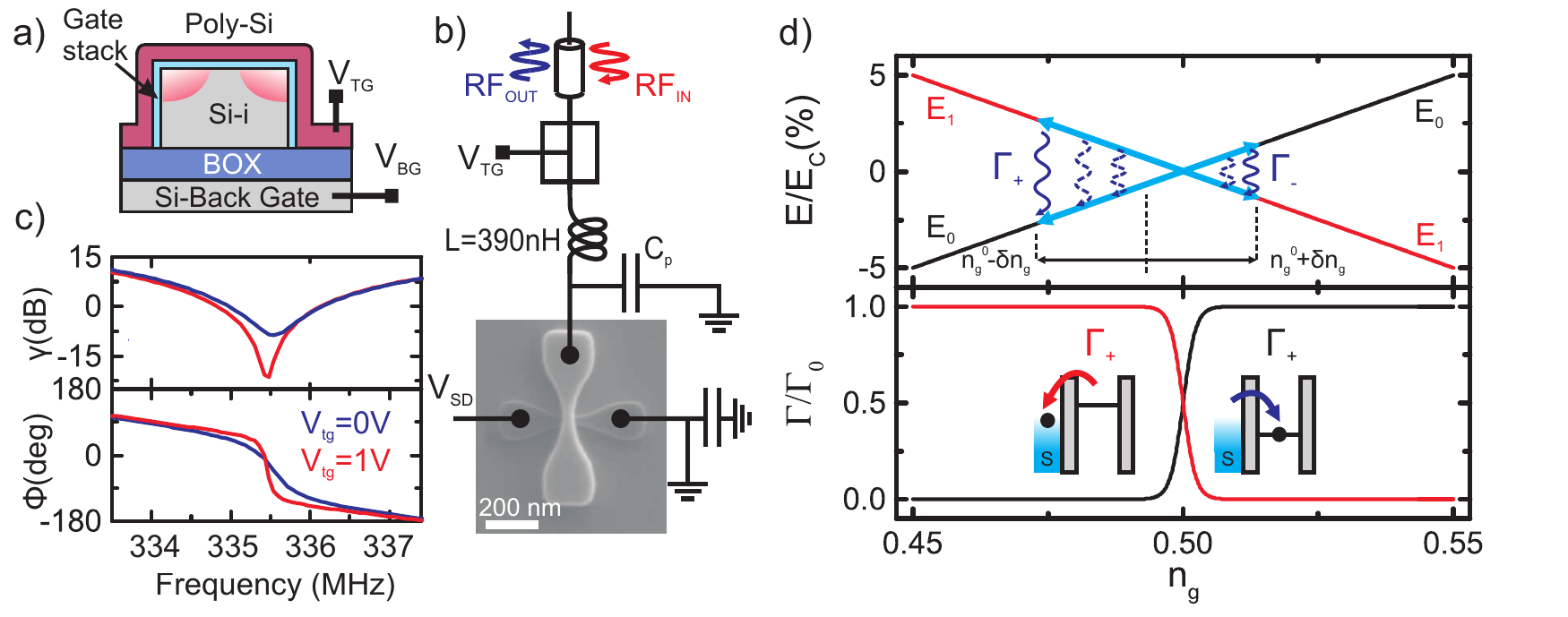}
	\caption{\textbf{Device and measurement set-up.} (a) Sketch of the cross-section of the device perpendicular to the transport direction. Due to the corner effect in silicon nanowire FETs and potential irregularities a parallel double quantum dot system forms when $V_{tg}$ is biased just below threshold. (b) Electron micrograph of an equivalent device ($l=64$~nm, $w=30$~nm) embedded in a resonant tank circuit. $C_p$ is the parasitic capacitance to ground and $L$ a surface mount inductor. $V_{sd}$,$V_{tg}$, and $V_{bg}$ are the DC voltages applied to the source, top gate, and back gate, respectively. (c) Characterization of the reflectometry response in magnitude (top-frame) and phase (bottom-frame) for the OFF ($V_{tg}=0$~V) and ON ($V_{tg}=1$~V) state of the transistor. (d) Top panel: Energy band diagram as a function of reduced gate voltage $n_g$. The initial detuning position is set by $n_g^0$ and $\delta n_g$ is the amplitude of the AC excitation. $\Gamma_{+ (-)}$ represents the tunnelling into (out of) the dot. Bottom panel: 3D-0D tunnel rates as a function of $n_g$ calculated for $E_C=15~meV$ and $T=100~mK$.}
	\label{Fig1}
\end{figure*}

The device used in this study is a narrow channel silicon-on-insulator nanowire transistors. A polycrystalline silicon top gate (length $l=64$~nm) wraps around three facets of the nanowire and is separated from the channel (width $w=30$~nm) by an oxide of equivalent thickness $1.3$~nm offering a strong control over the channel electrostatics. Due to the corner effect in silicon NWFETs~\cite{Sellier2007} electron accumulation happens first at the top most corners. Additionally, potential irregularities along the transport direction~\cite{Betz2014,Voisin2014} confine these corner channels creating a parallel double quantum dot system as schematically displayed in Fig.\ref{Fig1}(a). We couple the double-dot system to a resonant $LC$ circuit via the top gate electrode as depicted in Fig.\ref{Fig1}(b) and measure the complex polarisability using rf-reflectometry~\cite{Schoelkopf1998} and quadrature demodulation~\cite{Petersson2010}. At the resonant frequency ($f_r=1/2\pi\sqrt{LC_p}=334.8$~MHz, $C_p$ is the detector's stray capacitance) the magnitude ($\gamma$) and phase ($\phi$) components of the reflected signal are sensitive to admittance changes of the device (Fig.\ref{Fig1}(c)). Changes in the power dissipation in the system are captured in $\gamma$, whereas $\phi$ reflects susceptance changes such as tunneling or quantum capacitance~\cite{Ciccarelli2011, Colless2013,Chorley2012}. 

The origin of the gate-sensor signal can be understood in terms of electronic transitions in a fast-driven tunnel coupled two-level system~\cite{Persson2010a}. The detector is sensitive to the additional power that is dissipated when a charge is cyclically driven through a degeneracy point by an RF excitation with frequency ($f_0$) comparable to the tunnel rate. Moreover, a dispersive signal may be detected when electrons on average tunnel out-of-phase with rf cycle, generating an additional tunneling capacitance contribution $C_{t}$=$\alpha d\left\langle ne\right\rangle /dV_{tg}$ where $\left\langle ne\right\rangle$ is the average charge in the island. The levels $E_0$ (dot empty) and $E_1$ (dot full) are degenerate at $n_g=0.5$ as shown in Fig.\ref{Fig1}(d). The system is driven cyclically around a DC bias point $n_g^0=C_g V_g/e$ with amplitude $\delta n_g=C_g V_g^{rf}/e$, where $C_g$ is the gate capacitance and $V_g$ and $V_g^{rf}$ are the DC and RF gate voltage, respectively. Starting from the ground state at $n_g^0$, the fast RF excitation $\delta n_g$ moves the system non-adiabatically to the right of the degeneracy point. The system is now in the excited state until it relaxes. The excess dissipated energy $\Delta E=E_0-E_1$ is then captured as a change in the total reflected power of the device. The additional tunnelling capacitance term arises due to the fact that on average electron tunnelling leads the rf-excitation.

The mechanism has previously been studied in the context of Sisyphus dissipation for 3D metallic islands~\cite{Persson2010a,Ciccarelli2011}. Here, we transfer the concept to confined systems with a 0D density of states (DOS) coupled to 3D electron reservoirs. The corresponding tunnel rates are shown in the lower panel of Fig.\ref{Fig1}(d). A Fermi's golden rule calculation yields the general expression for 3D-0D tunnel rates,

\begin{equation}\label{Eq1}
	\Gamma_{\pm}=\frac{\Gamma_0}{1+e^{\pm \Delta E/k_B T}}
\end{equation}

where $k_B$ is the Boltzmann constant, $T$ the temperature, $\Delta E(t)$=$E_c(1-2n_g(t))$ the time-dependent energy difference and $n_g(t)=n_g^0+\delta n_g sin(2\pi f_0 t)$ is the normalized gate voltage. $\Gamma_0$ is the constant tunnel rate that establishes away from the degeneracy.

\begin{figure*}[htbp]
	\centering
		\includegraphics[width=1.0\textwidth]{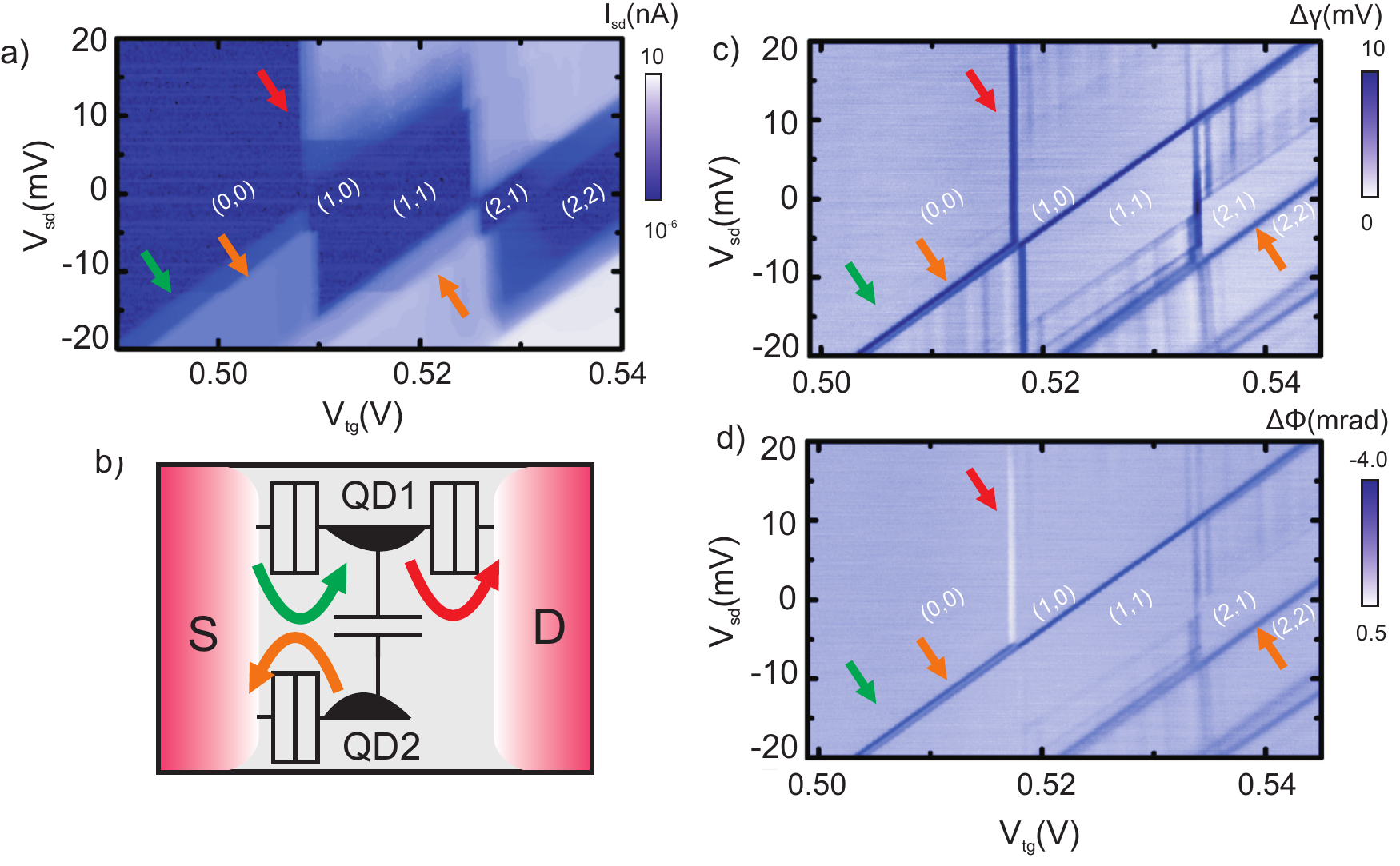}
	\caption{\textbf{DC / RF readout comparison}. Source-drain current $I_{sd}$ as a function of $V_{tg}$ and $V_{sd}$ for $V_{bg}$=2.4~V. The numbers (n,m) indicate the electron occupancy in the dots. The coloured arrows point to the onset of the transitions depicted in (b). (b) Schematic diagram of a top-view of the double-dot configuration. QD1 is tunnel coupled to the drain (source) indicated by the red (green) arrows while QD2 is only coupled to the source (orange arrow). (c) and (d) Gate sensor magnitude response $\Delta\gamma$ (c) and phase response $\Delta \phi$ in the same bias region as (a). Arrows indicate the same transitions as in (a).}
	\label{Fig2}
\end{figure*}

In order to evaluate the performance of the gate sensor we first compare the results to standard DC current measurements taken at 30~mK. Fig.\ref{Fig2}(a) shows the characteristic Coulomb diamonds where the electron occupancy is well defined. The onset of current happens at the diagonal lines marked by the green and red arrows indicating the alignment of the electrochemical levels of QD1 to the source and drain reservoirs, respectively (see Fig.\ref{Fig2}(b)). Moreover, we observe a shift of the high current regions at the orange lines, which can be understood as a capacitive shift of the QD1 transport characteristics due to the loading of an electron onto QD2. However, QD2 is only coupled to the source reservoir as depicted in Fig.\ref{Fig2}(b) and cannot be measured in transport. From this voltage shift we infer an equivalent charge sensing signal of $\Delta q/e=5.3\%$ and a mutual electrostatic energy of $0.58$~meV.

The magnitude response of the gate sensor in the same voltage region as the DC transport experiment is presented in Fig.\ref{Fig2}(c). The signal is enhanced at the edges of the charge stable regions in perfect match with the DC transport measurements. The detector maintains its sensitivity at high $V_{sd}$, as opposed to what is observed in 3D charge islands~\cite{Ciccarelli2011}. This additional advantage of our sensor stems from the independence of $\Gamma_\pm$ on $V_{sd}$ (Eq.1). Moreover, a striking difference between the DC and RF measurements is observed at the QD2 charge transitions (orange arrows): The magnitude of the gate sensor response is enhanced when the electrochemical levels of dot QD2 and the source reservoir are aligned demonstrating the possibility to detect charge transitions without the need of current flow or an external charge sensor.

In Fig.\ref{Fig2}(d) we extract the dispersive contributions by analyzing the phase response of the sensor. A change in the effective capacitance of the system due to electron tunnelling modifies the resonator's resonant frequency and hence causes a phase shift at the sampling frequency ($f_0$). The phase response relates to an effective change in the capacitance, $\Delta C$, of the system given by $\Delta\phi\approx-\pi Q\Delta C/C_p$, where Q is the quality factor of the resonator. We observe the response of the resonator as a phase shift $\Delta\phi$ at the edges of the conductive regions demonstrating dispersive readout of the charge state of a few-electron quantum dot system.  The sensor resolves phase changes of the order of 1~mrad which translates into capacitance detection of approximately 1~aF.  This phase resolution should allow quantum capacitance readout of strongly coupled quantum systems~\cite{Lambert2014}.

\begin{figure*}[htbp]
	\centering
		\includegraphics[width=0.95\textwidth]{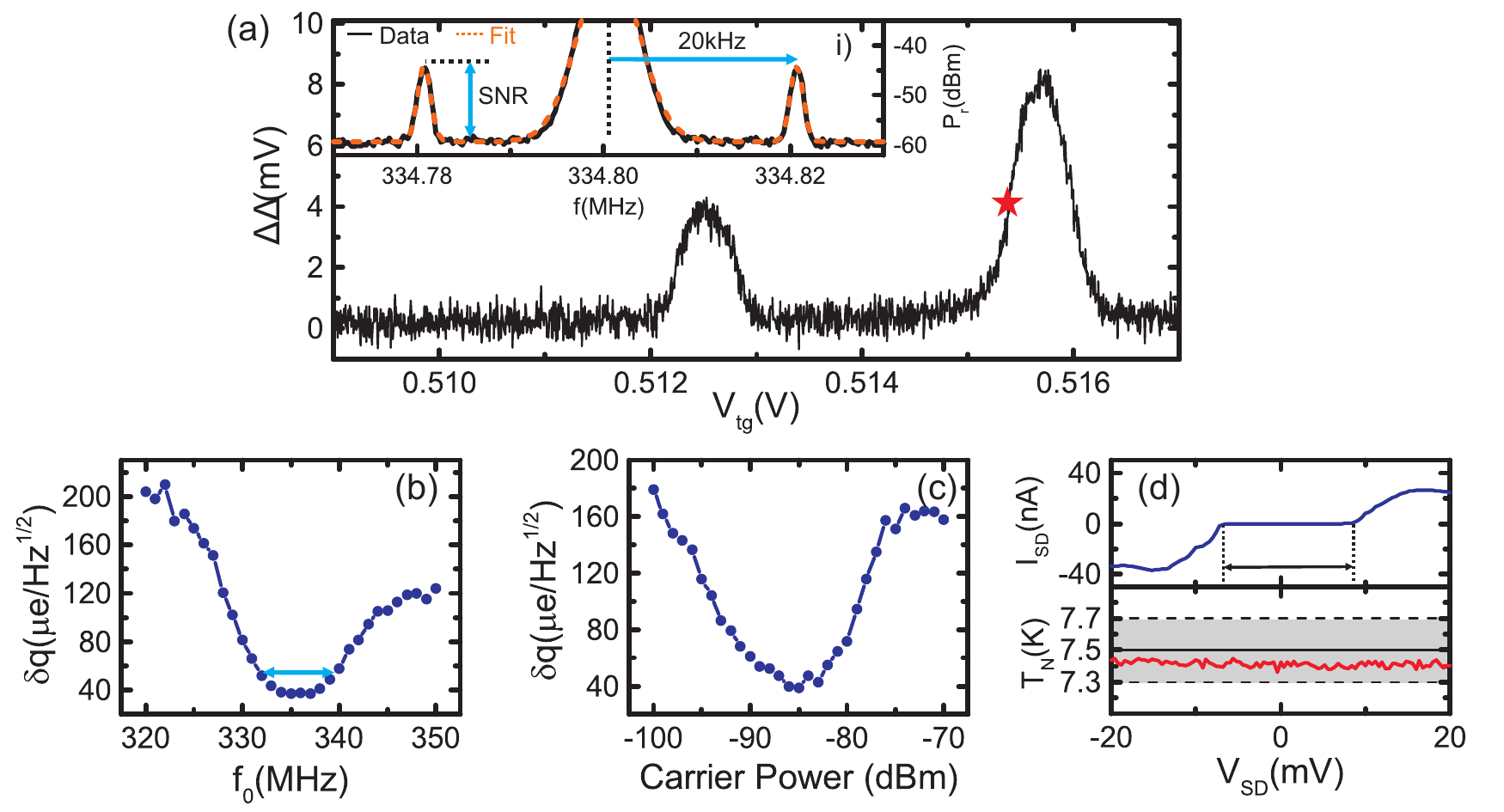}
	\caption{\textbf{Gate readout benchmark.} (a) Typical transfer curve $\Delta\gamma$-$V_{tg}$ showing two electron transitions. $V_{tg}$=0.5125~V is the first electron on QD1 and $V_{tg}$=0.5157~V is the first electron on QD2. Inset i) Sidebands at the point of maximum transconductance ($V_{tg}$=0.5155~V) at 20~kHz and equivalent excitation amplitude of 0.01e (solid black line) and fitted curve (dashed orange line). The measurement bandwidth is B=1~kHz. (b) and (c) Charge sensitivity as a function of carrier frequency (b) and carrier power (c). The 3~dB bandwidth, equivalent to 8~MHz, is indicated by the blue arrow. (d) Top panel: Source-drain current ($I_{SD}$) as a function of source-drain voltage $V_{SD}$ through a blockaded region at $V_{tg}$=0.58~V. Bottom panel: Noise temperature as a function of source-drain voltage measured at the resonant frequency with a resolution bandwidth of 1~MHz (red solid line). No rf excitation is applied during this measurement. The grey band indicates the calibrated noise temperature of the cryogenic amplifier measured using the same amplifier chain $T_N$=7.5$\pm$0.2~K.}
	\label{Fig3}
\end{figure*}

We now move on to the experimental characterization of the charge sensitivity of the gate sensor. In Fig.\ref{Fig3}(a) we show a typical transfer curve $\Delta\gamma$-$V_{tg}$ of the device comprising two charge transitions. We quantify the sensitivity by applying a small sinusoidal voltage (frequency $f_s$) to the top gate and monitoring the height of the sidebands in the frequency spectrum at $f_0\pm f_s$. We perform this characterization at the point of maximum transconductance of the QD2 transition (red star) since there $\Gamma_0$ and $f_0$ are well matched. The inset i shows the optimal sideband signal obtained with an equivalent voltage amplitude $\Delta q$=0.01e at $f_s$=20~kHz. This results in a signal-to-noise ratio (SNR) of 15.6~dB and a charge sensitivity of 37~$\mu$e/$\sqrt{Hz}$ given by $\Delta q/(\sqrt{2B}\times 10^{SNR/20})$~\cite{Nishiguchi2013}. It corresponds to an integration time $\tau\simeq 1.4$~ns required to resolve one electron charge and is comparable to typical charge coherence times in silicon and GaAs semiconductors~\cite{Shi2014,Petersson2010PRL}. Due to the strong capacitive coupling between the sensing electrode and the sensed element ($\alpha$=0.92) our detector shows a quantitative improvement of two orders of magnitude compared to recent GaAs gate sensors~\cite{Colless2013}. Additionally, we study the dependence of $\delta q$ as a function of the carrier frequency and carrier power. Fig.\ref{Fig3}(b) shows an optimal $\delta q$ for a carrier at 335~MHz with a detection bandwidth of 8~MHz (blue arrow) implying a cavity Q-factor of $42$. The optimal excitation power is found to be $-85$~dBm which is equivalent to $V_g^{rf}=0.5$~mV and an RF amplitude $\delta n_g=0.03$ (Fig.\ref{Fig3}(c)) 60 times larger larger than $k_B T/E_C$.

Ultimately, the sensitivity of rf-SETs is limited by shot noise due to the stochastic nature of the current through the tunnel barriers~\cite{Korotkov1994}. However the noise that ultimately limits gate-based charge sensors remains unknown. In Fig.\ref{Fig3}(d) we characterize the noise temperature of the system at the resonant frequency as a function of $V_{sd}$: The top panel shows the source-drain current across a Coulomb blockade region; the bottom panel displays the system's noise temperature, $T_N$ in the same region. $T_N$ remains constant independent of the device's current level and coincides with the noise temperature of the cryogenic amplifier, $7.5\pm 0.2$~K highlighted in grey~\footnote{The amplifier noise was calibrated using shot-noise thermometry~\cite{Spietz2003}}. We conclude that the noise, and hence also the sensitivity of the gated-based charge sensor presented here is experimentally limited by the noise level of the first cryogenic amplifying stage and is independent of $I_{sd}$ as opposed to shot-noise limited rf-SET detectors~\cite{Korotkov1994,Korotkov1999}.

The detection mechanism that governs gate-based readout is substantially different form rf-SETs where the largest contribution to the reflected power comes from the modulation of its differential conductance. In our case single-electron tunnelling is directly coupled to the frequency of the rf-drive and therefore a frequency-dependent noise spectrum is expected. To calculate the fundamental noise that limits gate-based charge detection we develop a model of the correlation of the power and phase fluctuations in a fast-driven two-level system. 

We first focus on the dissipative components associated to fast electronic transitions. We present in the following a numerical calculation of the instantaneous dissipated power, $P(t)$, due to the rf excitation, the corresponding spectral density of the fluctuations in dissipated power, $S_{pp}$, and the charge sensitivity, $\delta q$, using experimental parameters ($E_c$, $T$, $f_0$) obtained earlier.

\begin{figure*}[htbp]
	\centering
		\includegraphics{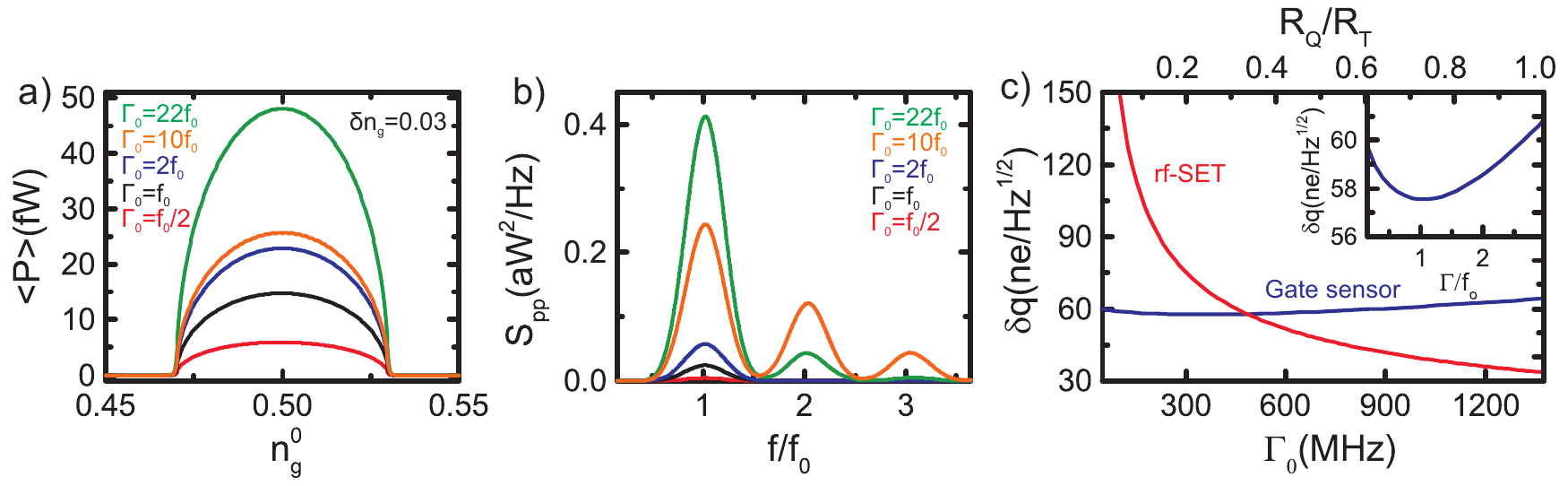}
	\caption{\textbf{Sisyphus noise spectrum and ultimate charge sensitivity.} (a) Calculated average power dissipation as a function of reduced gate voltage $n_g^0$ for different tunnel rates $\Gamma_0$ and reduced rf amplitude $\delta n_g$=0.03. (b) Power spectral density of the power noise, $S_{pp}$, as a function of normalized frequency $f/f_0$ for several tunnel rates at $n_g^0=n_g^{0,max}$. (c) Gate-based noise-limited charge sensitivity as a function of tunnel rate (blue line). rf-SET ultimate sensitivity as a function of the ratio between the quantum resistance and the tunnel resistance $R_Q/R_T$ (red trace) from~\cite{Korotkov1999}. $R_T=2k_BT/e^2\Gamma_0$ at $\Delta E$=0. Inset: Zoom-in close to the resonant frequency. The best sensitivity is achieved at the resonant frequency $f_0$. $T$=100~mK and $E_C$=15~meV throughout the calculations.}
	\label{Fig5}
\end{figure*}

The dynamics of power dissipation are given by a master equation~\cite{Persson2010a}. For the 0D-3D tunnel rates (Eq.\eqref{Eq1}) it reduces to solving the differential equation,

\begin{equation}\label{Eq_diffEqP1}
\dot P_1(t)+\Gamma_0 P_1(t)=\Gamma_+(t)
\end{equation} 

where $P_1$ is the probability of the electron being in the dot. Solving Eq.\eqref{Eq_diffEqP1} for $P_1(t)$ we obtain the instantaneous dissipated power

\begin{equation}
	P(t)=\Delta E(t)[\Gamma_0 P_1(t)+\Gamma_+(t)]
\end{equation}

as well as its average, $\left\langle P\right\rangle$, over one period $f_0^{-1}$, which is shown in Fig.\ref{Fig5}(a) as a function of DC offset $n_g^0$ for several tunnel rates, using the optimal RF excitation $\delta n_g=0.03$. The most power is dissipated when the system is biased at the degeneracy and no dissipation occurs for $\left|n_g^0-0.5\right|\gtrsim\delta n_g$, as expected. The change in average power dissipation, $d\left\langle P\right\rangle/dn_g^0$, is maximum at $n_g^{0,max}\approx n_g^0\pm \delta n_g $. 

With respect to tunnel rates, $\left\langle P\right\rangle$ peaks at $\Gamma_0=2\pi f_0$ showing a way to maximize the detector response. This can be understood from the analytical approximation given in Eq.\eqref{Eq_P_analyt}, where $\Gamma_{\pm}$ is expanded around $n_g^0$ to the first order of $\delta n_g$.

\begin{equation} \label{Eq_P_analyt}
\left\langle P\right\rangle\simeq \frac{(eV_g^{rf}\alpha)^2}{k_BT}\frac{1}{1+cosh\left(\frac{\Delta E(0)}{k_B T}\right)}\frac{\Gamma_0}{1+\Gamma_0^2/\omega_0^2}
\end{equation}

Note that this approximation is only valid for excitations $\delta n_g\ll k_B T/E_c$. The numerical solution $P_1(t)$ of Eq.\eqref{Eq_diffEqP1} furthermore allows us to calculate the power fluctuations around the mean value and subsequently the power spectral density of the power noise, $S_{pp}$ (see Methods). The results are shown in Fig.\ref{Fig5}(b) as a function of reduced frequency $f/f_0$ for several $\Gamma_0$ at $n_g^0=n_g^{0,max}$. At the point of maximum sensitivity the power noise is maximum at frequencies that match the excitation frequency $f_0$ and its harmonics. As a whole these calculations show a new type of frequency-periodic noise, the Sisyphus noise, in which the stochastic nature of electron tunnelling is directly coupled to the resonator's natural frequency of oscillation. However, signal and noise frequency decoupling could be achieved at the degeneracy point (See Supplementary).

Knowledge of the Sisyphus noise and the change in average power dissipation allows us now to calculate the charge sensitivity

\begin{equation} \label{Eq_dq}
\delta q = \frac{\sqrt{S_{pp}(\omega)}}{d\left\langle P\right\rangle/d\,n_g^0}
\end{equation} 

in units of $e/\sqrt{Hz}$. The result is presented in Fig.\ref{Fig5}(c) (main panel and inset): In blue we show the charge sensitivity as a function of tunnel rate for $S_{pp}(\omega)$=$S_{pp}(2\pi f_0)$ and $n_g^0=n_g^{0,max}$, which corresponds to the experimental situation presented earlier. For our experimental setup and a realistic range of tunnel rates the fundamental limit of dissipative rf charge read-out is $\delta q\lesssim 70~ne/\sqrt{Hz}$. As can be seen from the inset of Fig.\ref{Fig5}(c), the best charge sensitivity occurs for matching tunnel rate and excitation frequency $f_0$. Adjusting $\Gamma_0$ and $f_0$ may thus be a route to obtain the best sensitivity, though experimental constraints such as the resonant circuit, i.e. $f_0$, will set a practical limit.

We benchmark our detector's ultimate sensitivity against that of rf-SETs predicted in Ref.~\cite{Korotkov1999} (red trace in Fig.\ref{Fig5}(c)). A direct comparison is made considering the tunnel resistance equivalent $R_T=2k_BT/e^2\Gamma_0$, which sets the same tunnel rate at $\Delta E$=0 for both cases. Comparing both curves, we find that rf-SETs exhibit a better fundamental charge sensitivity at high tunnel rates, whereas our rf gate sensor performs favourably at low $\Gamma_0$, i.e. for highly resistive tunnel barriers which is the usual scenario of quantum dots in the few-electron regime.

We now move on to the analysis of dispersive readout and the limits of gate-based phase sensing. The change in the probability of the electron being in the dot due to the rf-drive generates a tunnelling capacitance term given by,

\begin{equation} \label{Eq_Phi}
C_{t}(t)=e\alpha\frac{dP_1}{dV_g}=\frac{e\alpha}{V_g^{rf}cos(\omega_0t)}\dot P_1(t)
\end{equation}

\begin{figure*}[htbp]
	\centering
		\includegraphics{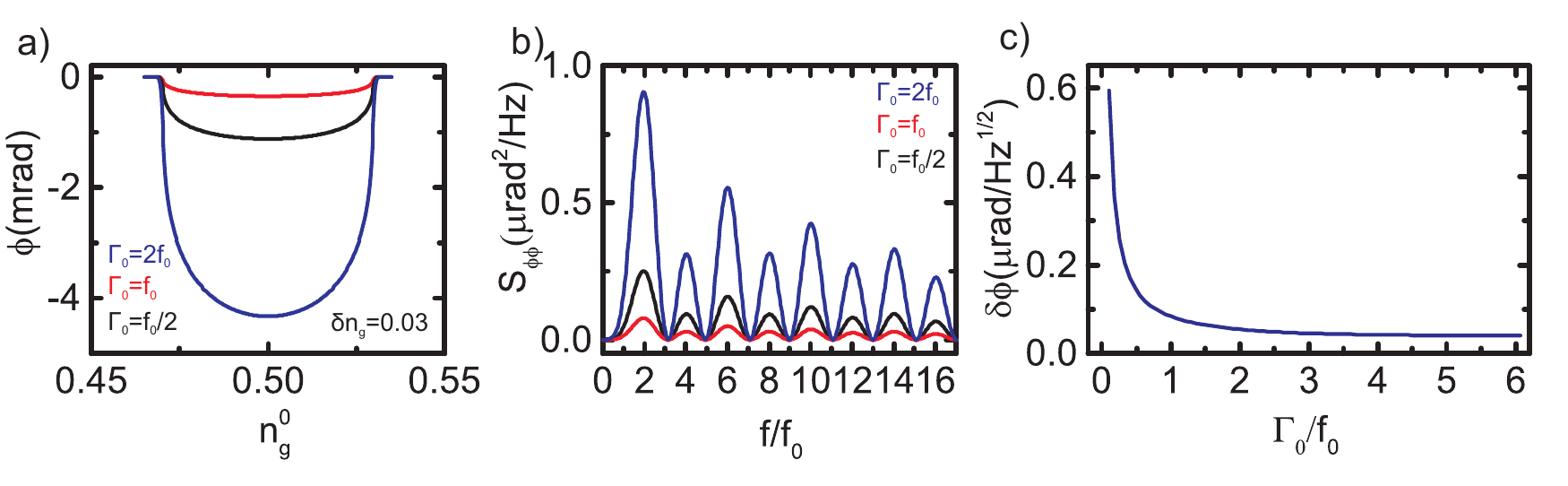}
	\caption{\textbf{Phase noise and sensitivity.}(a) Calculated average phase shift as a function of reduced gate voltage $n_g^0$ for different tunnel rates $\Gamma_0$ and reduced rf amplitude $\delta n_g$=0.03. (b) Power spectral density of the phase noise, $S_{\phi\phi}$, as a function of normalized frequency $f/f_0$ for several tunnel rates. (c) Noise-limited phase sensitivity as a function of reduced tunnel rate. For all calculations $T$=100~mK and $E_C$=15~meV. }
	\label{Fig6}
\end{figure*}

The associated average phase shift as a function of the offset $n_g^0$ is plotted in Fig.\ref{Fig6}(a). Similarly to the average power dissipation the phase response peaks at the degeneracy point and the change $d\left\langle\phi\right\rangle/dn_g^0$ is maximum close to $n_g^{0,max}$. The tunnelling capacitance is always positive independently of temperature and tunnel rate and hence the phase shift remains negative. This results indicates that on average electron tunnelling leads the rf excitation unless the transient behaviour of the system is probed (see Supplementary). An analytical solution of Eq.\eqref{Eq_Phi} for $\delta n_g\ll k_BTE_C$ confirms this result,

\begin{equation} \label{Eq_Phi_analyt}
\left\langle\Delta\phi\right\rangle\simeq -\pi Q\frac{e\alpha}{k_BTC_p}\frac{1}{1+cosh\left(\frac{\Delta E(0)}{k_B T}\right)}\frac{1}{1+\omega_0^2/\Gamma_0^2}
\end{equation} 

In contrast to the average power dissipation, the phase shift does not peak at the resonant frequency and tends to a constant value for large tunnelling rates. This offers a way to maximise the dispersive signal while reducing the excess power dissipation.

The phase noise spectral density associated with the phase fluctuations is presented in Fig.\ref{Fig6}(b). It is maximum at multiple of $2f_0$ with odd harmonics predominantly contributing to the noise spectrum. This can be explained from the time symmetry of the tunnelling capacitance: $C_t\approx tan(\omega_0t)$ is an odd function of time and periodic in $2f_0$ (see Supplementary). Signal and noise are hence decoupled in frequency space offering an opportunity for high-sensitivity phase detection.

The detector's ultimate phase sensitivity, $\delta \phi$, is calculated in the same way as its charge counterpart (see Eq.\eqref{Eq_dq}) as the ratio of phase noise $\sqrt{S_{\phi\phi}}$ and change in average phase $d\left\langle\phi\right\rangle/d\,n_g^0$.
It is plotted in Fig.\ref{Fig6}(c) as a function of reduced tunnel rate at bias $n_g^{0,max}$ and frequency $\omega = 2\pi f_0$. In tune with $\left\langle \phi\right\rangle$ the phase sensitivity has no minimum at $f_0$ but approaches a low constant value at high tunnel rate. 
For parameters corresponding to our setup the fundamental limit of phase sensitivity is $\delta \phi \lesssim 0.04~\mu rad/\sqrt{Hz}$ for tunnel rates $\Gamma_0>>f_0$, which translates into an integration time of $\sim 0.16~ps$. 
Fast gate-based phase detection is thus most appropriate for systems with transparent tunnel barriers, but performs well even at moderate tunnel rates. 
In comparison, phase detection based on cooper-pair transistors provides sensitivities of the order of $1~\mu rad/\sqrt{Hz}$~\cite{Roschier2005}, whereas resonant detection applied to the source of a nanowire yields phase detection times $\tau_{min}\sim 9~\mu s$~\cite{Jung2012}. 

In conclusion, we have reported the high sensitivity charge readout of an interacting few-electron double quantum dot system. Our results show that gate-based readout can offer a competitive alternative to the best rf-SET charge sensors at an equivalent measurement bandwidth. Moreover, the flexibility of this technique, which does not require additional external sensor elements, opens up a window for high-fidelity, high-integration qubit architectures. Future work may explore quantum capacitance readout of interdot transitions and excited-state microwave manipulation.

%
\section{Acknowledgment}
The authors thank D.A. Williams for fruitful discussion. The research leading to these results has been supported by the European Community's seventh Framework under the Grant Agreement No. 318397. The samples presented in this work were designed and fabricated by the TOLOP project partners, http://www.tolop.eu.

\section{Additional information}
The authors declare no competing financial interests.\newline\mbox{}\newline
A.C.B. and M.F.G. contributed equally to this work. A.C.B. and M.F.G. devised and performed the experiments; S.B. fabricated the sample; A.C.B. performed the calculations; M.F.G. and A.C.B. co-wrote the paper; A.J.F. contributed with the theoretical background.

\section{Methods}
\paragraph{\textbf{Device fabrication}}\mbox{}\newline
The nanowire transistors used in this study were fabricated on silicon-on-insulator (SOI) substrates with a 145~nm buried oxide. The silicon layer is patterned to create the nanowires by means of optical lithography, followed by a resist trimming process. For the gate stack, 1.9~nm HfSiON capped by 5~nm TiN and 50~nm polycrystalline silicon were deposited leading to a total equivalent oxide thickness (EOT) of 1.3~nm. The Si thickness under the HfSiON/TiN gate is 11~nm. After gate etching, a SiN layer (thickness 10~nm) was deposited and etched to form a first spacer on the sidewalls of the gate. 18~nm-thick Si raised source and drain contacts were selectively grown prior to the source/drain extension implantation and activation annealing. Then a second spacer was formed and followed by source/drain implantations, activation spike anneal, and salicidation (NiPtSi).
\paragraph{\textbf{Measurement setup}}\mbox{}\newline
Measurements were performed at the base temperature of a dilution refrigerator with an electron temperature of 100~mK. Radio-frequency reflectometry was performed at 335~MHz by embedding the sample in an resonant $LC$ circuit formed by a surface mount inductor (390~nH) and the device's parasitic capacitance to ground (500~fF). After low-temperature and room temperature amplification the reflected signal was fed into a quadrature demodulator, yielding the signal's quadrature and in-phase component from which magnitude $\gamma$ and phase $\Phi$ are calculated.
\paragraph{\textbf{Numerical calculations}}\mbox{}\newline
$P_1(t)$ and all subsequent derivatives such as $P(t)$ or $\Delta \phi(t)$ are finite time series in the numerical calculations. $P_1(t)$ was obtained by solving Eq.\eqref{Eq_diffEqP1} numerically for a range of parameters corresponding to experimental constraints, namely $f_0$=335~MHz, $T$=100~mK, $E_C$=15~meV, and $\delta n_g$=0.03, at equal time steps in the interval $t=[0,5f_0^{-1}]$. The averages $\left\langle P\right\rangle$ and $\left\langle \Delta\phi\right\rangle$ are taken over one full period. Due to the finite time series nature of all variables no auto-correlation is necessary, but $S_{pp}$ ($S_{\phi\phi}$) can be calculated immediately from the power (phase) fluctuations: $S_{XX}(\omega)=(\Delta t)^2 f_0 \left| \sum_n (X(n \Delta t) - \left\langle X\right\rangle)\times e^{-i\omega n}\right|^2$.

\end{document}